\documentclass{article} % For LaTeX2e
\usepackage[final]{colm2026_conference}

\usepackage{amsmath}
\usepackage{microtype}
\usepackage{hyperref}
\usepackage{url}
\usepackage{booktabs}
\usepackage{tcolorbox}
\tcbuselibrary{breakable}
\usepackage{lineno}

\definecolor{darkblue}{rgb}{0, 0, 0.5}
\hypersetup{colorlinks=true, citecolor=darkblue, linkcolor=darkblue, urlcolor=darkblue}

\title{Human-Likeness is Multi-Dimensional: Evaluating LLM-Generated Virtual Patients Against Simulated Patient Dialogue}

\author{
\begin{tabular}{c}
\textbf{Reyhaneh Hosseinpourkhoshkbari}$^{1}$\thanks{Corresponding author: \texttt{reyhaneh.hosseinpour@utdallas.edu}},
\textbf{Wei-Chen Huang}$^{1}$,
\textbf{Yile Wang}$^{1}$, \\[2pt]
\textbf{Alireza Irani Helabad}$^{1}$,
\textbf{Zhiyu Chen}$^{2}$,
\textbf{Richard M. Golden}$^{1}$ \\[4pt]
{\small\normalfont $^{1}$School of Behavioral and Brain Sciences, The University of Texas at Dallas} \\
{\small\normalfont $^{2}$Department of Computer Science, The University of Texas at Dallas}
\end{tabular}
}

\begin{document}
\raggedbottom
\ifcolmsubmission
\linenumbers
\fi
\maketitle
\begin{abstract}
Large language models (LLMs) are increasingly used to generate virtual patients (VPs) for medical training. As these systems move toward broader deployment in medical education, reliable evaluation is needed to identify potential failure modes and ensure that generated patient responses remain consistent with human simulated patient (SP) behavior. Inaccurate, unrealistic, or inconsistent responses may introduce educational risks by exposing learners to misleading patient interactions. However, current evaluation approaches rely heavily on human judgment, making large-scale evaluation difficult.
In this work, we present a controlled generation pipeline for systematically comparing VP responses generated by 10 LLMs across multiple decoding temperatures with human respiratory SP responses. To support scalable evaluation, we introduce a multi-dimensional framework for measuring similarity between LLM-generated and human SP dialogue across semantic, conversational grounding, linguistic, and emotional dimensions.
Across semantic, conversational, linguistic, and emotional dimensions, no single model consistently produced the most human-like responses. From a practical perspective, Gemini 2.5 Flash demonstrated the most consistently human-like performance across semantic, linguistic, grammatical, and emotional dimensions, while Claude Sonnet 4.6 achieved the strongest semantic fidelity. These findings highlight the importance of multi-dimensional evaluation and demonstrate how automated metrics can provide a scalable approach for assessing virtual patient systems before broader deployment in medical education.

\end{abstract}

\section{Introduction}

Clinical communication skills are a core component of medical education because effective doctor–patient interactions are essential for accurate diagnosis, treatment planning, patient satisfaction, and healthcare outcomes \citep{hosseinpourkhoshkbari2025clinical, aper2015should,stewart1995effective,boissy2016communication}. Poor communication has been linked to medical errors, particularly during history taking, and may contribute to increased healthcare costs and reduced quality of care \citep{rogers2006analysis,singh2013types,ahmadipour2015medical}. As a result, providing safe and effective communication training remains a major goal of medical education.

Simulated environments play an important role in achieving this goal by allowing learners to practice patient interactions in controlled settings without exposing real patients to unnecessary risk \citep{ten2018approaches}. Simulated patients (SPs), who are human actors who act as patients, provide highly realistic training experiences but are costly and time-intensive to recruit, train, and maintain \citep{barrows1993overview,mavis2006developments,johnson2020impact}. Virtual patients (VPs) offer a scalable alternative that can support flexible, repeatable, and accessible clinical training.

Recent advances in large language models (LLMs) have led to a new generation of conversational virtual patients capable of producing natural and interactive dialogue \citep{cook2024creating}. Prior work has demonstrated promising results using a variety of prompting approaches and evaluation methods \citep{holderried2024generative,borg2024enhancing,sardesai2024utilizing,aster2025chatgpt,yi2025feasibility,yuan2025ai}. However, LLM-generated virtual patients remain susceptible to problems such as hallucinations, character inconsistency, unrealistic behavior, and inaccurate clinical details \citep{holderried2024generative,sardesai2024utilizing,borg2024enhancing,grevisse2024raspatient,ji2023survey}. Because these systems may eventually be deployed at scale in medical education, understanding their reliability is important. Inaccurate or unrealistic patient responses may reduce educational quality and expose learners to misleading patient interactions, creating potential downstream healthcare risks.

A second challenge concerns evaluation. Most existing studies rely on expert judgment, learner feedback, or task-specific assessments. While valuable, these approaches are often expensive, difficult to scale, and provide limited insight into model-specific strengths, weaknesses, and potential failure modes. As a result, it remains difficult to systematically compare models or monitor the quality of generated patient dialogue across different systems and settings.

Third, prompt design varies widely, from minimal inputs to highly structured scenarios, with limited guidance on achieving consistency and realism \citep{holderried2024generative,aster2025chatgpt}. This variability complicates comparisons across studies and makes it difficult to identify effective design choices.

Together, these challenges highlight the need for scalable evaluation methods that can support the pre-screening of dialogue resemblance prior to the deployment of LLM-generated virtual patients. Such methods should move beyond subjective judgments and evaluate how closely generated responses resemble human patient communication across multiple dimensions. In this work, we use simulated patient dialogue as a practical reference because real patient conversations are often difficult to access due to privacy constraints. Although simulated patient dialogue is not a perfect gold standard, it provides a realistic and widely used proxy for human patient interaction in medical training.

\section{Related work}
Recent studies have explored the use of LLMs for virtual patient generation using a variety of prompting strategies. \citet{holderried2024generative} employed structured illness-script prompts and evaluated generated patients using measures such as medical plausibility and script consistency. \citet{aster2025chatgpt} used symptom-focused prompts and assessed realism through user experience measures. \citet{borg2024enhancing} incorporated detailed patient descriptions and conversational context, reporting improved learner experiences but also identifying challenges such as response delays and repetition. Other studies have evaluated LLM-generated virtual patients using expert ratings of relevance, accuracy, fluency, and behavioral realism \citep{yi2025feasibility} or through usability-focused evaluations \citep{yuan2025ai}.

Overall, prior work demonstrates that LLMs can support realistic and engaging virtual patient interactions. However, most evaluations rely on subjective assessments and focus on a limited number of models, scenarios, or evaluation criteria. Relatively little work has systematically compared multiple LLMs using a common framework or examined similarity to human-simulated patient dialogue across semantic, conversational, linguistic, and emotional dimensions.
\subsection*{Aims}
This study examines how closely VP responses generated by different LLMs resemble human simulated patient dialogue. To do so, we present a controlled generation pipeline for comparing 10 LLMs across multiple decoding temperatures and construct a synthetic respiratory VP dialogue corpus by re-generating human SP responses. We then introduce a multi-dimensional evaluation framework that measures similarity to human SP dialogue across semantic, conversational, linguistic, and emotional dimensions. Using this framework, we identify model-specific strengths and weaknesses and show that no single model consistently performs best across all dimensions. These findings highlight the importance of evaluating VP systems using multiple complementary measures rather than a single notion of realism.

\section{Methods}
%%%%%%%%%%%%%%%%%%%%%%%%%%%%%%%%%%%%%%%%%%%%%%%%

%%%%%%%%%%%%%%%%%%%%%%%%%%%%%%%%%%%%%%%%%%%%%%%%%%
\paragraph{Dataset.}
The dataset used in this study consists of 21 respiratory-case transcripts extracted from \citet{fareez2022dataset}, comprising 1,095 paired clinician–SP dialogue turns. Because real patient conversational data is often unavailable due to privacy constraints, SP dialogue serves as a practical human reference for simulated clinical communication.

\paragraph{Large language models.}
Let $\mathcal{M}={m_1,\ldots,m_K}$ denote the set of LLMs evaluated in this study. We evaluated 10 models from multiple providers, including OpenAI (GPT-5.2, GPT-4o), Google (Gemini-2.5-Flash-Preview, Gemini-3.1-Flash-Lite-Preview), Anthropic (Claude-Sonnet-4.6, Claude-Haiku-4.5), Meta LLaMA (LLaMA-3.3-70B-Instruct-Turbo, LLaMA-3.1-8B-Instruct), and Mistral (Mixtral-8x7B-Instruct-v0.1, Mistral-Small-24B-Instruct-2501). Table~\ref{tab:llm_models} provides additional details. The selected models represent a range of model families, sizes, and capabilities, enabling a broad comparison of LLM-based virtual patient generation.

\paragraph{Prompt construction.}
For each transcript, we constructed an illness script using a structured template containing symptoms, medical history, social history, and other clinically relevant information (Appendix~\ref{appendix:appendixa}, Box~\ref{box:example_script}).The template was designed to organize the clinically relevant patient information available in the source cases into standard history-taking categories. Human annotators extracted this information from the original clinician--SP conversations to create structured patient profiles that remained grounded in realistic clinical cases.

\paragraph{Transcript generation.}
Let $\mathcal{T}={0.0,0.5,1.0}$ denote the set of generation temperatures. For each case $i$, model $m_k \in \mathcal{M}$, and temperature $\tau \in \mathcal{T}$, a VP response $y_{i,k,\tau}$ was generated from the corresponding patient prompt and dialogue context:

\begin{equation}
y_{i,k,\tau} \sim m_k(s_i, c_{i,t}, r_{i,t}; \tau),
\end{equation}

where $s_i$ denotes the structured patient prompt, $c_{i,t}$ denotes the recent dialogue context at turn $t$, and $r_{i,t}$ denotes the running patient state. Smaller values of $\tau$ reduce response variability, whereas larger values increase variability \citep{renze2024effect}.

% \begin{figure*}[t]
%     \centering
%     \includegraphics[width=\textwidth]{vp_generation_figure.pdf}
%     \caption{
%     Controlled turn-by-turn virtual patient (VP) generation framework.
%     At each turn, a clinician utterance from the reference transcript is
%     combined with a structured patient profile, a running patient state
%     updated every four clinician turns, and the 10 most recent dialogue
%     messages. These inputs are provided to each LLM at temperature
%     $\tau \in \{0.0,0.5,1.0\}$ to generate a VP response, which is appended
%     to the dialogue history for the subsequent turn.
%     }
%     \label{fig:vp_generation}
% \end{figure*}
To generate complete VP transcripts, clinician utterances from the original conversations were provided sequentially to each model. This controlled setup was intentionally used to ensure that each model responded to the same clinician utterances as the human SPs, enabling direct turn-level comparisons while minimizing variation introduced by differences in clinician behavior. At each turn, the model received: (1) a structured patient profile, (2) a running patient state, and (3) a sliding window containing the most recent dialogue history. The running state was initialized from the patient profile and updated periodically during the interaction to maintain consistency regarding demographics, symptom history, disclosed and undisclosed information, key clinical findings, and affective characteristics. In our implementation, the dialogue window contained the 10 most recent messages and the patient state was refreshed every four clinician turns. This combination provided access to recent conversational context while maintaining a compact representation of longer-term patient information. Generation settings were otherwise standardized across all models to ensure fair comparison.

\section{Evaluation framework}

Our goal is to evaluate how closely LLM-generated VP responses resemble human SP dialogue. We therefore use an automated evaluation framework that measures similarity to human SP dialogue across four complementary dimensions: semantic similarity, conversational grounding, linguistic characteristics, and emotional expression. 

Rather than interpreting individual metrics as direct measures of educational quality or clinical correctness, we use them to compare how closely generated responses resemble patterns observed in human SP dialogue. Our framework is therefore intended as a scalable and reproducible screening tool that complements, rather than replaces, expert evaluation.

\subsection{Semantic similarity}

Semantic similarity measures whether generated responses convey information similar to that expressed by human SPs, independent of differences in wording. This is important in clinical dialogue because different phrasings may communicate the same symptoms or contextual information, whereas changes in meaning can alter clinical interpretation \citep{hosseinpourkhoshkbarican, chang1998importance,ten2017principles,cuffy2020measuring}.

We compute embedding-based similarity using BioClinicalBERT and BGE-base-en-v1.5. BioClinicalBERT provides a clinically specialized representation, while BGE provides a general semantic baseline. For each human response $h$ and generated response $y$, we compute cosine similarity, Manhattan similarity, and Pearson correlation between their embeddings:

\begin{align}
\mathrm{Sim}_{\mathrm{cos}}(h, y)
&= \frac{\phi(h)^\top \phi(y)}
{\|\phi(h)\|_2\,\|\phi(y)\|_2}
&&\in [-1,1], \label{cos} \\
\mathrm{Sim}_{\mathrm{Manhattan}}(h, y)
&= -\|\phi(h)-\phi(y)\|_1
&&\in (-\infty,0], \label{Manhattan} \\
\mathrm{Sim}_{\mathrm{pearson}}(h, y)
&= \frac{(\phi(h)-\mu_h)^\top(\phi(y)-\mu_y)}
{\|\phi(h)-\mu_h\|_2\,\|\phi(y)-\mu_y\|_2}
&&\in [-1,1], \label{pearson}
\end{align}
where $\mu_h=\mathbb{E}[\phi(h)]$ and $\mu_y=\mathbb{E}[\phi(y)]$ denote the means of the embedding vectors.

\subsection{Conversational grounding}

We examine conversational grounding, defined as the degree to which a patient response remains semantically related to the immediately preceding clinician utterance. In clinical dialogue, patient responses are expected to remain connected to the physician's questions and prompts. We operationalize this construct using embedding-based similarity between each clinician utterance and the corresponding patient response.

For each turn $t$, let $x_t$ denote the clinician utterance and $y_t$ denote the patient response. We compute:

\begin{equation}
G_t = \mathrm{Sim}\big(\phi(x_t), \phi(y_t)\big),
\end{equation}

where $\phi(\cdot)$ denotes the embedding function and $\mathrm{Sim}(\cdot,\cdot)$ denotes one of the similarity measures defined in equations \ref{cos} to \ref{pearson}.

We compute grounding scores using both BioClinicalBERT and BGE embeddings. Our goal is to compare the conversational grounding of LLM-generated VP responses with that observed in human SP dialogue under the same clinician utterances. These scores are therefore interpreted relative to the human SP reference distribution rather than as absolute measures of response quality or conversational appropriateness.

\subsection{Linguistic characteristics}

Semantic similarity does not capture differences in language use, style, or grammatical structure. We therefore examine linguistic characteristics using both lexical diversity and part-of-speech (POS) distributions.

Lexical diversity is measured using word count, corrected type-token ratio (CTTR), measure of textual lexical diversity (MTLD), and Yule's $K$, computed using the \texttt{LexicalRichness} library \citep{lex}. These measures quantify vocabulary diversity and lexical concentration while accounting for differences in response length \citep{hosseinpourkhoshkbarican,                                  tweedie1998variable,malvern2004lexical,mccarthy2010mtld,covington2010cutting}.

To characterize grammatical structure, we also analyze POS distributions across standardized grammatical categories, including nouns, verbs, adjectives, adverbs, and related classes \citep{cutting1992practical,kumawat2015pos,divyapushpalakshmi2021efficient}. Comparing these distributions allows us to assess how closely generated language resembles human SP language beyond semantic content alone.

\subsection{Emotional expression}

Emotional expression is an important component of patient communication that is not fully captured by semantic or linguistic measures. Patients communicate not only information about their condition but also emotional states that influence the interaction \citep{guo2023close,heyman2020patient,beach2021clinician}.

To evaluate emotional patterns, we use the pretrained {j-hartmann/emotion-english-distilroberta-base} model \citep{hartmann2022emotion}, which classifies text into emotion categories. We then compare the distribution of predicted emotions in generated VP responses with those observed in human SP dialogue to assess how closely models reproduce human-like emotional expression.

\section{Results}

We address three research questions: (1) How closely do LLM-generated virtual patient responses resemble human simulated patient dialogue? (2) How do models differ across semantic, conversational, linguistic, and emotional dimensions? (3) Does any single model consistently produce the most human-like responses?
\subsection{Semantic similarity to human SP dialogue}
 A Friedman test revealed significant differences among models under both BioClinicalBERT ($\chi^2(9)=1509.72$, $p<.001$) and BGE ($\chi^2(9)=1569.22$, $p<.001$). 

\begin{figure*}[t]
\centering
\includegraphics[width=1\textwidth]{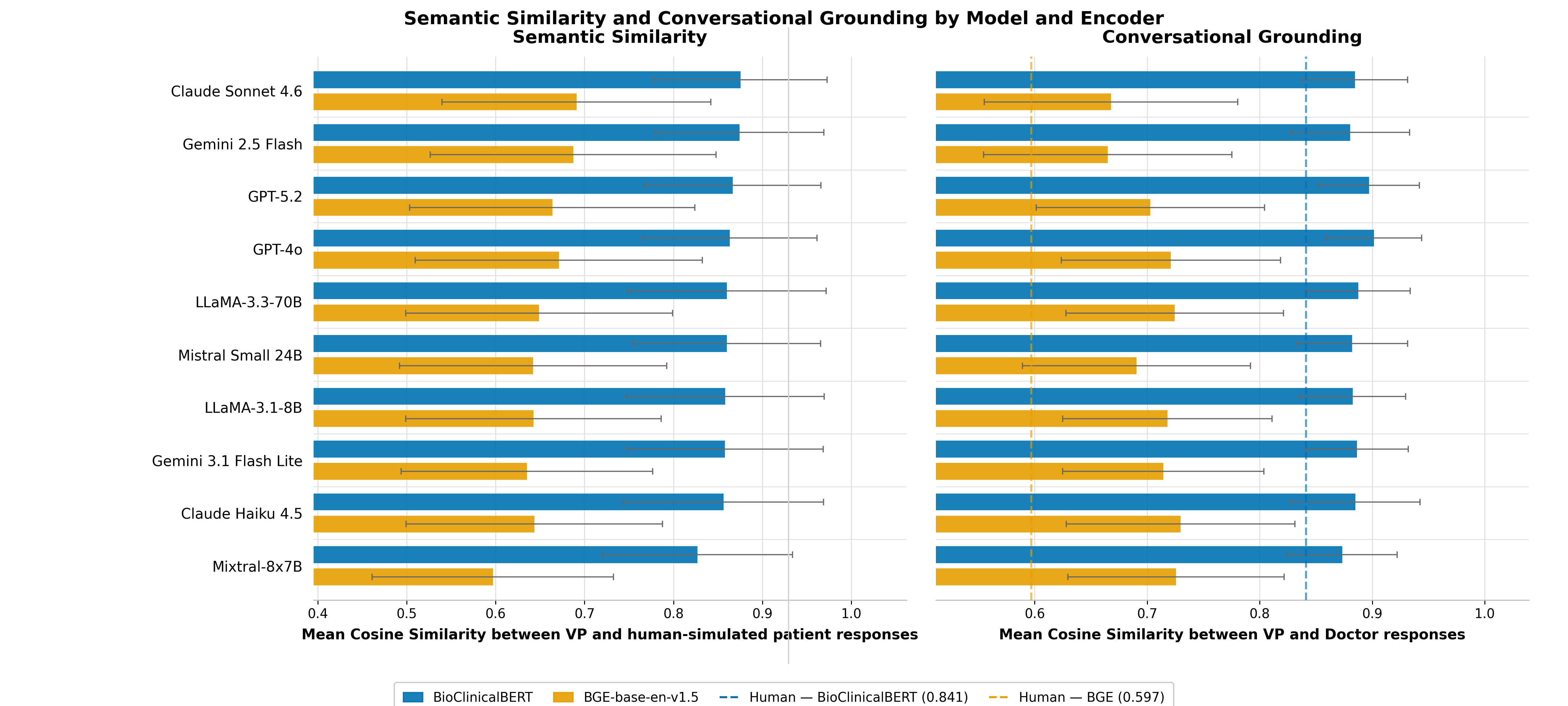}
\caption{
Semantic similarity (left) and conversational grounding (right) across models. Bars show mean cosine similarity using BioClinicalBERT and BGE embeddings. Dashed lines indicate human SP grounding baselines.
}
\label{fig:semantic_grounding}
\end{figure*}

As shown in Figure~\ref{fig:semantic_grounding} (left), model rankings were highly consistent across encoders. Claude Sonnet 4.6 and Gemini 2.5 Flash achieved the highest semantic similarity to human SP responses, followed by GPT-5.2 and GPT-4o. In contrast, Mixtral-8x7B consistently produced the lowest similarity scores and was significantly different from all other models, with medium-to-large effect sizes ($|r|=0.27$--$0.68$). While several pairwise differences among the remaining models reached statistical significance, effect sizes were generally small, indicating substantial overlap among the middle and upper-performing models.

Temperature had negligible impact on similarity scores or model rankings. Overall, the results suggest that most evaluated models reproduce human SP content with broadly comparable fidelity, with Claude Sonnet 4.6 and Gemini 2.5 Flash forming the highest-performing group and Mixtral-8x7B constituting a distinct lower-performing tier.
\subsection{Conversational grounding}

A Friedman test revealed significant differences among models under both BioClinicalBERT ($\chi^2(9)=1017.22$, $p<.001$) and BGE ($\chi^2(9)=1030.51$, $p<.001$). 

As shown in Figure~\ref{fig:semantic_grounding} (right), all models exceeded the human SP grounding baseline under both encoders. Under BioClinicalBERT, the human baseline was 0.841, whereas model scores ranged from 0.874 to 0.902, with Mixtral-8x7B showing the smallest deviation from the human baseline. Under BGE, the human baseline was 0.597 and all models again exceeded it, with scores ranging from 0.665 to 0.730 and Gemini 2.5 Flash showing the smallest deviation from the human baseline. This consistent pattern suggests that LLM-generated responses are more tightly coupled to the immediately preceding clinician utterance than human SP responses.

Unlike semantic similarity, conversational grounding produced substantially different patterns across models and encoders. Because all model grounding scores exceeded the human SP baseline, lower model scores corresponded to smaller deviations from human dialogue. Mixtral-8x7B showed the smallest deviation under BioClinicalBERT, whereas Gemini 2.5 Flash showed the smallest deviation under BGE.

These results suggest that conversational grounding captures a distinct aspect of dialogue behavior from semantic similarity. Models that more closely reproduce human patient content do not necessarily exhibit the strongest clinician-response associations. 

\subsection{Linguistic characteristics}

\paragraph{Lexical analysis} Friedman tests revealed significant model differences across all metrics (words: $\chi^2(9)=6778.02$; terms: $\chi^2(9)=6747.79$; CTTR: $\chi^2(9)=6295.40$; MTLD: $\chi^2(9)=3329.32$; Yule's $K$: $\chi^2(9)=2624.18$; all $p<.001$).

The biggest differences were observed in response length. Human SPs produced relatively short responses (12.9 words per turn), and GPT-4o (11.8 words) and Gemini 2.5 Flash (13.2 words) were the closest to this baseline. In contrast, several models generated substantially longer responses, including Gemini 3.1 Flash Lite (41.3 words), LLaMA-3.3-70B (38.9 words), and Claude Haiku 4.5 (36.0 words). Mixtral-8x7B was an extreme outlier, averaging 89.7 words per turn, nearly seven times the human baseline.

Lexical diversity analyses revealed an opposite pattern. Human SP responses exhibited lower lexical diversity than all evaluated models, reflecting the constrained and repetitive vocabulary characteristic of natural patient speech. GPT-4o and Gemini 2.5 Flash were closest to the human baseline on both MTLD and CTTR, whereas Gemini 3.1 Flash Lite and Mixtral-8x7B produced the most lexically diverse responses. Vocabulary repetition measured by Yule's $K$ showed that Claude Sonnet 4.6 most closely matched the human SP distribution, while GPT-4o and GPT-5.2 exhibited substantially lower repetition and more varied vocabulary.

\begin{figure}[t]
\centering
\includegraphics[width=0.75\linewidth]{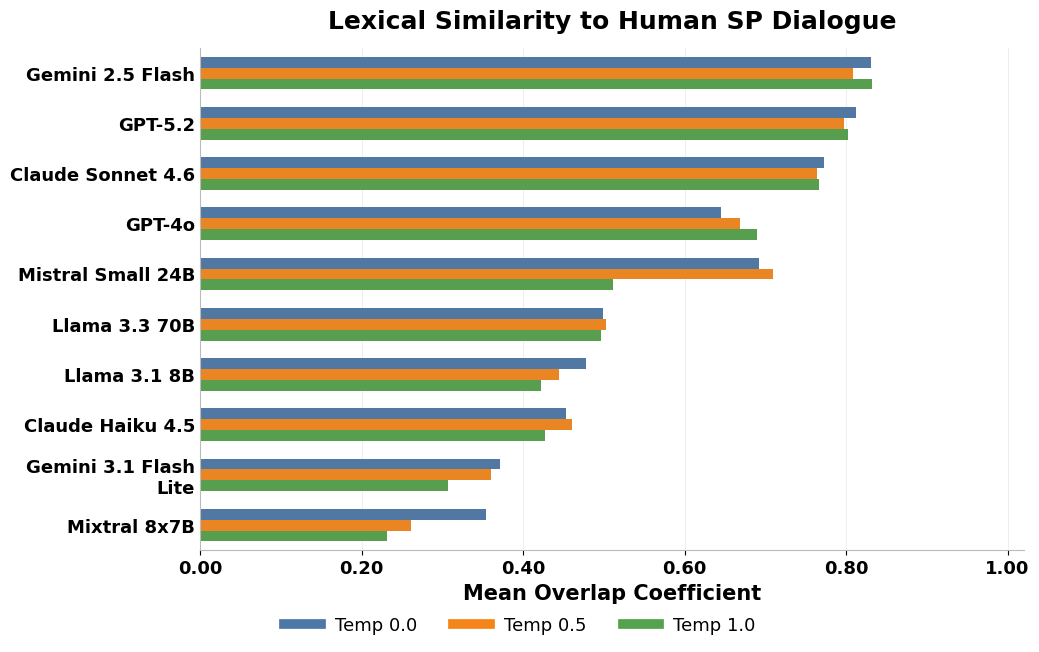}
\caption{
Mean lexical similarity to human SP dialogue was measured using overlap coefficients across three temperature settings. Higher values indicate closer agreement with human lexical distributions. Rankings were largely stable across temperatures, with Gemini 2.5 Flash, GPT-5.2, and Claude Sonnet 4.6 showing the highest similarity and Mixtral-8x7B showing the lowest.
}
\label{fig:lexical_similarity}
\end{figure}

Overall, lexical results reveal a length-diversity trade-off. GPT-4o and Gemini 2.5 Flash most closely matched human response length but produced vocabulary that was more diverse than human SP dialogue. In contrast, Claude Sonnet 4.6 more closely reproduced human vocabulary repetition patterns but generated longer responses. No single model consistently matched human SP dialogue across all lexical dimensions. However, Gemini 2.5 Flash and GPT 5.2 showed the highest similarity to human SP on average (Figure \ref{fig:lexical_similarity}).

\paragraph{Part-of-Speech distributions}

We compared part-of-speech (POS) distributions between model-generated and human SP responses using 13 grammatical categories. 
Gemini 2.5 Flash produced the most human-like POS distributions, differing significantly from human SPs on only 6 of 13 tags. GPT-4o ranked second, with significant differences on 9 of 13 tags. In contrast, all remaining models differed significantly from human SP dialogue on 12--13 of 13 tags, indicating substantially weaker alignment with human grammatical patterns.

A consistent finding across all models was the underproduction of interjections (INTJ), including fillers, discourse markers, and hesitation expressions that commonly occur in natural patient speech. Human SPs produced an average of 1.12 interjections per turn, whereas all models generated significantly fewer ($p<10^{-7}$). Claude Sonnet 4.6 (0.96) and Claude Haiku 4.5 (0.89) were closest to the human baseline, while Mistral Small 24B (0.48) showed the largest deviation. More broadly, no model successfully matched human distributions for determiners (DET), interjections (INTJ), nouns (NOUN), or particles (PART), with significant differences observed across all models.

Mixtral-8x7B emerged as a notable outlier, exhibiting the largest deviations from human POS distributions. The strongest effects were observed for nouns, pronouns, and verbs, largely reflecting the model's substantially longer responses relative to human SP dialogue. In contrast, temperature had minimal influence on POS patterns, with significance patterns remaining highly stable across all three temperature settings.

Overall, POS analyses suggest that while Gemini 2.5 Flash and GPT-4o most closely approximate human grammatical structure, all evaluated models exhibit systematic differences from human SP language, particularly in their reduced use of conversational markers such as interjections and fillers.

\subsection{Emotional expression}

Human SP responses were predominantly neutral (69.3\%), followed by disgust (13.5\%) and sadness (8.5\%), with relatively low fear (2.6\%) and anger (0.6\%).

Omnibus G-tests showed that emotion distributions differed significantly between generated and human SP responses across all model settings (all $p<10^{-94}$), with small but consistent effects (Cramér's $V=0.128$--$0.142$). Compared with human SPs, LLM-based VP produced neutral responses by 18.3 percentage points and over-produced negative emotions, particularly fear (+6.9 pp), disgust (+6.9 pp), and sadness (+5.3 pp).

Model-level deviations varied substantially. Gemini 2.5 Flash produced the closest overall emotion distribution to human SPs, with the lowest mean Jensen-Shannon divergence (JS = 0.090), whereas Mixtral-8x7B showed the largest deviation (JS = 0.249). These results suggest that LLM-generated patients tend to express more negative affect than human SPs, indicating an affective bias that may influence perceived realism in virtual patient interactions (see Table \ref{tab:emotion_deviation}).

\begin{table}[t]
\centering
\caption{Temperature-aggregated emotion-wise deviation from human emotion distributions. Smaller values indicate closer similarity to human emotion proportions.}
\label{tab:emotion_deviation}
\small
\begin{tabular}{lcccc}
\toprule
\textbf{Model} & \textbf{Neutral} & \textbf{Disgust} & \textbf{Sadness} & \textbf{Fear} \\
\midrule
Claude Sonnet 4.6      & \textbf{0.1068} & 0.0837 & 0.0051 & \textbf{0.0264} \\
Gemini 2.5 Flash       & 0.1167 & 0.0393 & 0.0484 & 0.0290 \\
Claude Haiku 4.5       & 0.1291 & 0.0762 & \textbf{0.0022} & 0.0539 \\
GPT-5.2                & 0.1831 & 0.1093 & 0.0423 & 0.0316 \\
Llama 3.3 70B          & 0.2066 & 0.0725 & 0.0370 & 0.0971 \\
Llama 3.1 8B           & 0.2167 & 0.0906 & 0.0564 & 0.0697 \\
GPT-4o                 & 0.2232 & 0.1055 & 0.0791 & 0.0387 \\
Gemini 3.1 Flash Lite  & 0.2352 & \textbf{0.0374} & 0.0867 & 0.1111 \\
Mistral Small 24B      & 0.2656 & 0.0682 & 0.0899 & 0.1075 \\
Mixtral 8x7B           & 0.2775 & 0.0473 & 0.0675 & 0.1627 \\
\bottomrule
\end{tabular}
\end{table}

\section{Discussion}
This study examined how closely VP responses generated by different LLMs resemble human SP dialogue across semantic, conversational, linguistic, and emotional dimensions.

Our results demonstrate that similarity to human simulated patient dialogue is inherently multi-dimensional. Models that performed well on one dimension often performed less well on others. For example, Claude Sonnet 4.6 and Gemini 2.5 Flash achieved high semantic similarity to human SP responses, whereas the models varied in their deviation from the human conversational grounding baseline. Likewise, GPT-4o and Gemini 2.5 Flash most closely matched human response length, while Claude Sonnet 4.6 more closely reproduced human vocabulary repetition patterns. These findings indicate that no single model consistently produced the most human-like behavior across all evaluated dimensions, highlighting the limitations of relying on a single evaluation metric.

Several systematic differences between LLM-generated and human SP dialogue also emerged. Most models produced responses that were more tightly coupled to clinician utterances than human SPs, suggesting a tendency toward over-grounding. Models generally generated more lexically diverse language than human SPs and consistently under-produced interjections, fillers, and hesitation markers that characterize natural patient speech. In addition, emotion analyses revealed a tendency toward increased negative affect relative to human SPs. Together, these findings suggest that current LLM-based virtual patients can reproduce clinical content with relatively high fidelity, but still differ from human SPs in conversational style, linguistic behavior, and emotional expression.

From a practical perspective, Gemini 2.5 Flash showed relatively human-like patterns across several evaluated dimensions, while Claude Sonnet 4.6 achieved the strongest semantic fidelity, further illustrating that no single model consistently aligned most closely with human SPs across all measures. More broadly, our results show that automated multidimensional evaluation can identify strengths and weaknesses that would be missed by a single realism measure. Such evaluation may provide a scalable screening approach for assessing VP systems before more costly human evaluation and broader deployment in medical education.

This study has several limitations. First, our study was conducted using respiratory simulated-patient transcripts from a single dataset and may not generalize to other clinical domains or patient populations. Second, our evaluation focused on text-based dialogue and did not assess clinical accuracy, educational effectiveness, or multimodal interactions. Therefore, resemblance to human SP dialogue should not be interpreted as evidence of clinical correctness or educational suitability. Third, the automated metrics used in this study were not validated against judgments from clinicians, simulated-patient trainers, or learners. Future work should examine whether these computational measures correspond to human perceptions of VP realism and human-likeness and whether they are associated with educational outcomes. We will also extend the framework to broader clinical settings and investigate how automated evaluation can be combined with expert assessment to support reliable deployment of virtual patient systems.

\bibliography{colm2026_conference}

@article{mavis2006developments,
  title={Developments: faculty, students, and actors as standardized patients: expanding opportunities for performance assessment},
  author={Mavis, Brian and Turner, Jane and Lovell, Kathryn and Wagner, Dianne},
  journal={Teaching and learning in medicine},
  volume={18},
  number={2},
  pages={130--136},
  year={2006},
  publisher={Taylor \& Francis}
}

@article{ten2018approaches,
  title={Approaches to assessing the clinical reasoning of preclinical students},
  author={Cate, Th. J. ten and Durning, Steven J},
  journal={Principles and practice of case-based clinical reasoning education: a method for Preclinical Students},
  pages={65--72},
  year={2018},
  publisher={Springer International Publishing}
}

@article{johnson2020impact,
  title={Impact of standardized patients on first semester nursing students self-confidence, satisfaction, and communication in a simulated clinical case},
  author={Johnson, Kelly V and Scott, Allison L and Franks, Lisa},
  journal={SAGE open nursing},
  volume={6},
  pages={2377960820930153},
  year={2020},
  publisher={SAGE Publications Sage CA: Los Angeles, CA}
}

@article{cook2024creating,
  title={Creating virtual patients using large language models: scalable, global, and low cost},
  author={Cook, David A},
  journal={Medical teacher},
  pages={1--3},
  year={2024},
  publisher={Taylor \& Francis}
}

@article{boissy2016communication,
  title={Communication skills training for physicians improves patient satisfaction},
  author={Boissy, Adrienne and Windover, Amy K and Bokar, Dan and Karafa, Matthew and Neuendorf, Katie and Frankel, Richard M and Merlino, James and Rothberg, Michael B},
  journal={Journal of general internal medicine},
  volume={31},
  pages={755--761},
  year={2016},
  publisher={Springer}
}

@article{stewart1995effective,
  title={Effective physician-patient communication and health outcomes: a review.},
  author={Stewart, Moira A},
  journal={CMAJ: Canadian medical association journal},
  volume={152},
  number={9},
  pages={1423},
  year={1995},
  publisher={Canadian Medical Association}
}

@article{ahmadipour2015medical,
  title={Medical error and related factors during internship and residency},
  author={Ahmadipour, Habibeh and Mortazavi,Nahid},
  journal={Indian J Med Ethics},
  volume={12},
  number={4},
  pages={215--9},
  year={2015}
}

@article{singh2013types,
  title={Types and origins of diagnostic errors in primary care settings},
  author={Singh, Hardeep and Giardina, Traber Davis and Meyer, Ashley ND and Forjuoh, Samuel N and Reis, Michael D and Thomas, Eric J},
  journal={JAMA internal medicine},
  volume={173},
  number={6},
  pages={418--425},
  year={2013},
  publisher={American Medical Association}
}

@article{rogers2006analysis,
  title={Analysis of surgical errors in closed malpractice claims at 4 liability insurers},
  author={Rogers Jr, Selwyn O and Gawande, Atul A and Kwaan, Mary and Puopolo, Ann Louise and Yoon, Catherine and Brennan, Troyen A and Studdert, David M},
  journal={Surgery},
  volume={140},
  number={1},
  pages={25--33},
  year={2006},
  publisher={Elsevier}
}

@article{barrows1993overview,
  title={An overview of the uses of standardized patients for teaching and evaluating clinical skills. AAMC},
  author={Barrows, Howard S},
  journal={Academic medicine},
  volume={68},
  number={6},
  pages={443--51},
  year={1993},
  publisher={LWW}
}

@article{aper2015should,
  title={“Should I prioritize medical problem solving or attentive listening?”: The dilemmas and challenges that medical students experience when learning to conduct consultations},
  author={Aper, Leen and Veldhuijzen, Wemke and Dornan, Tim and van de Ridder, Monica and Koole, Sebastiaan and Derese, Anselme and Reniers, Jan},
  journal={Patient education and counseling},
  volume={98},
  number={1},
  pages={77--84},
  year={2015},
  publisher={Elsevier}
}

@article{fareez2022dataset,
  title={A dataset of simulated patient-physician medical interviews with a focus on respiratory cases},
  author={Fareez, Faiha and Parikh, Tishya and Wavell, Christopher and Shahab, Saba and Chevalier, Meghan and Good, Scott and De Blasi, Isabella and Rhouma, Rafik and McMahon, Christopher and Lam, Jean-Paul and others},
  journal={Scientific Data},
  volume={9},
  number={1},
  pages={313},
  year={2022},
  publisher={Nature Publishing Group UK London}
}

@article{holderried2024generative,
  title={A generative pretrained transformer (GPT)--powered chatbot as a simulated patient to practice history taking: prospective, mixed methods study},
  author={Holderried, Friederike and Stegemann--Philipps, Christian and Herschbach, Lea and Moldt, Julia-Astrid and Nevins, Andrew and Griewatz, Jan and Holderried, Martin and Herrmann-Werner, Anne and Festl-Wietek, Teresa and Mahling, Moritz and others},
  journal={JMIR medical education},
  volume={10},
  number={1},
  pages={e53961},
  year={2024},
  publisher={JMIR Publications Inc., Toronto, Canada}
}

@article{ji2023survey,
  title={Survey of hallucination in natural language generation},
  author={Ji, Ziwei and Lee, Nayeon and Frieske, Rita and Yu, Tiezheng and Su, Dan and Xu, Yan and Ishii, Etsuko and Bang, Ye Jin and Madotto, Andrea and Fung, Pascale},
  journal={ACM computing surveys},
  volume={55},
  number={12},
  pages={1--38},
  year={2023},
  publisher={ACM New York, NY}
}

@article{aster2025chatgpt,
  title={ChatGPT as a Virtual patient: written empathic expressions during medical history taking},
  author={Aster, Alexandra and Ragaller, Sophia Viktoria and Raupach, Tobias and Marx, Ambra},
  journal={Medical Science Educator},
  pages={1--10},
  year={2025},
  publisher={Springer}
}

@article{yuan2025ai,
  title={AI agent as a simulated patient for history-taking training in clinical clerkship: an example in stomatology},
  author={Yuan, Yongxiang and He, Jieyu and Wang, Fang and Li, Yaping and Guan, Chaxiang and Jiang, Canhua},
  journal={Global Medical Education},
  year={2025},
  publisher={De Gruyter}
}

@inproceedings{grevisse2024raspatient,
  title={RasPatient Pi: A Low-Cost Customizable LLM-Based Virtual Standardized Patient Simulator},
  author={Gr{\'e}visse, Christian},
  booktitle={International Conference on Applied Informatics},
  pages={125--137},
  year={2024},
  organization={Springer}
}

@article{borg2024enhancing,
  title={Enhancing clinical reasoning skills for medical students: a qualitative comparison of LLM-powered social robotic versus computer-based virtual patients within rheumatology},
  author={Borg, Alexander and Jobs, Benjamin and Huss, Viking and Gentline, Cidem and Espinosa, Fabricio and Ruiz, Mini and Edelbring, Samuel and Georg, Carina and Skantze, Gabriel and Parodis, Ioannis},
  journal={Rheumatology International},
  volume={44},
  number={12},
  pages={3041--3051},
  year={2024},
  publisher={Springer}
}

@article{sardesai2024utilizing,
  title={Utilizing generative conversational artificial intelligence to create simulated patient encounters: a pilot study for anaesthesia training},
  author={Sardesai, Neil and Russo, Paolo and Martin, Jonathan and Sardesai, Anand},
  journal={Postgraduate medical journal},
  volume={100},
  number={1182},
  pages={237--241},
  year={2024},
  publisher={Oxford University Press}
}

@article{yi2025feasibility,
  title={The feasibility of using generative artificial intelligence for history taking in virtual patients},
  author={Yi, Yongjin and Kim, Kyong-Jee},
  journal={BMC Research Notes},
  volume={18},
  number={1},
  pages={80},
  year={2025},
  publisher={Springer}
}

@misc{lex,
        author = {Shen, Lucas},
        doi = {10.5281/zenodo.6607007},
        license = {MIT license},
        title = {{LexicalRichness: A small module to compute textual lexical richness}},
        url = {https://github.com/LSYS/lexicalrichness},
        year = {2022}
}

@article{covington2010cutting,
  title={Cutting the Gordian knot: The moving-average type--token ratio (MATTR)},
  author={Covington, Michael A and McFall, Joe D},
  journal={Journal of quantitative linguistics},
  volume={17},
  number={2},
  pages={94--100},
  year={2010},
  publisher={Taylor \& Francis}
}

@article{mccarthy2010mtld,
  title={MTLD, vocd-D, and HD-D: A validation study of sophisticated approaches to lexical diversity assessment},
  author={McCarthy, Philip M and Jarvis, Scott},
  journal={Behavior research methods},
  volume={42},
  number={2},
  pages={381--392},
  year={2010},
  publisher={Springer}
}

@article{tweedie1998variable,
  title={How variable may a constant be? Measures of lexical richness in perspective},
  author={Tweedie, Fiona J and Baayen, R Harald},
  journal={Computers and the Humanities},
  volume={32},
  pages={323--352},
  year={1998},
  publisher={Springer}
}

@book{malvern2004lexical,
  title={Lexical diversity and language development},
  author={Malvern, David and Richards, Brian and Chipere, Ngoni and Dur{\'a}n, Pilar},
  year={2004},
  publisher={Springer}
}

@inproceedings{cutting1992practical,
  title={A practical part-of-speech tagger},
  author={Cutting, Douglass and Kupiec, Julian and Pedersen, Jan and Sibun, Penelope},
  booktitle={Third conference on applied natural language processing},
  pages={133--140},
  year={1992}
}

@article{kumawat2015pos,
  title={POS tagging approaches: A comparison},
  author={Kumawat, Deepika and Jain, Vinesh},
  journal={International Journal of Computer Applications},
  volume={118},
  number={6},
  year={2015},
  publisher={Citeseer}
}

@article{divyapushpalakshmi2021efficient,
  title={An efficient sentimental analysis using hybrid deep learning and optimization technique for Twitter using parts of speech (POS) tagging},
  author={Divyapushpalakshmi, M and Ramalakshmi, R},
  journal={International Journal of Speech Technology},
  volume={24},
  number={2},
  pages={329--339},
  year={2021},
  publisher={Springer}
}

@article{hosseinpourkhoshkbari2025clinical,
  title={Clinical Reasoning Assessment Methods: Current Simulated Environment Practice and Future Prospects using {AI} and Psychometrics},
  author={Hosseinpourkhoshkbari, Reyhaneh and Golden, Richard M.},
  journal={SSRN Electronic Journal},
  year={2025},
  doi={10.2139/ssrn.5400917}
}

@inproceedings{hosseinpourkhoshkbarican,
  title={Can You Spot the Virtual Patient ({VP})? Expert Evaluation, Turing Test, Linguistic Analysis, and Semantic Similarity Analysis},
  author={Hosseinpourkhoshkbari, Reyhaneh and Huang, Wei-chen and Muttreja, Suvel and Golden, Richard M.},
booktitle={NeurIPS 2025 Workshop: The Second Workshop on GenAI for Health: Potential, Trust, and Policy Compliance},
year={2025}
}

@article{guo2023close,
  title={How close is chatgpt to human experts? comparison corpus, evaluation, and detection},
  author={Guo, Biyang and Zhang, Xin and Wang, Ziyuan and Jiang, Minqi and Nie, Jinran and Ding, Yuxuan and Yue, Jianwei and Wu, Yupeng},
  journal={arXiv preprint arXiv:2301.07597},
  year={2023}
}

@article{ten2017principles,
  title={Principles and practice of case-based clinical reasoning education: a method for preclinical students},
  author={Ten Cate, Olle and Custers, Eug{\`e}ne JFM and Durning, Steven J},
  year={2017},
  publisher={Springer}
}

@article{chang1998importance,
  title={The importance of early problem representation during case presentations},
  author={Chang, ROWLAND W and Bordage, Georges and Connell, Karen J},
  journal={Academic medicine: journal of the Association of American Medical Colleges},
  volume={73},
  number={10 Suppl},
  pages={S109--S111},
  year={1998}
}

@article{cuffy2020measuring,
  title={Measuring the quality of patient--physician communication},
  author={Cuffy, Clint and Hagiwara, Nao and Vrana, Scott and McInnes, Bridget T},
  journal={Journal of biomedical informatics},
  volume={112},
  pages={103589},
  year={2020},
  publisher={Elsevier}
}

@article{beach2021clinician,
  title={Clinician response to patient emotion: impact on subsequent communication and visit length},
  author={Beach, Mary Catherine and Park, Jenny and Han, Dingfen and Evans, Christopher and Moore, Richard D and Saha, Somnath},
  journal={The Annals of Family Medicine},
  volume={19},
  number={6},
  pages={515--520},
  year={2021},
  publisher={The Annals of Family Medicine}
}

@article{heyman2020patient,
  title={Patient experience and expression of unpleasant emotions during health care encounters},
  author={Heyman, Richard E and Baucom, Katherine JW and Giresi, Jill and Isaac, Liza J and Slep, Amy M Smith},
  journal={Journal of Patient Experience},
  volume={7},
  number={6},
  pages={969--972},
  year={2020},
  publisher={SAGE Publications Sage CA: Los Angeles, CA}
}

@inproceedings{renze2024effect,
  title={The effect of sampling temperature on problem solving in large language models},
  author={Renze, Matthew},
  booktitle={Findings of the association for computational linguistics: EMNLP 2024},
  pages={7346--7356},
  year={2024}
}

@misc{hartmann2022emotion,
  title={Emotion english distilroberta-base},
  author={Hartmann, Jochen},
  year={2022}
}
\bibliographystyle{colm2026_conference}
\newpage
\appendix
\section{Appendix}
\label{appendix:appendixa}

\begin{table}[t]
\centering
\caption{Large Language Models Evaluated}
\label{tab:llm_models}
\small
\setlength{\tabcolsep}{4pt}
\begin{tabular}{p{0.58\linewidth} p{0.38\linewidth}}
\toprule
\textbf{Model Family \& Variants} & \textbf{Source} \\
\midrule

OpenAI (GPT-5.2, GPT-4o) 
& \url{openai.com} \\

Google (Gemini-3-Flash, Gemini-3.1-Flash-Lite) 
& \url{deepmind.google/technologies/gemini} \\

Anthropic (Claude-Sonnet-4.6, Claude-Haiku-4.5) 
& \url{anthropic.com/claude} \\

Meta LLaMA (LLaMA-3.3-70B, LLaMA-3.1-8B) 
& \url{ai.meta.com/llama} \\

Mistral (Mixtral-8x7B, Mistral-Small-24B) 
& \url{mistral.ai/models} \\

\bottomrule
\end{tabular}
\end{table}

\newcounter{mybox}
\begin{tcolorbox}[
  title={Box \refstepcounter{mybox}\label{box:example_script}\themybox: Excerpt from an Example Ilness Script},
  colback=gray!10,
  colframe=gray!70!black,
  colbacktitle=gray!70!black,
  coltitle=white,
  fonttitle=\bfseries,
  boxrule=0.4pt,
  sharp corners=south,
  before skip=6pt, after skip=6pt,
  left=4pt, right=4pt, top=4pt, bottom=4pt,
  width=\textwidth,
  breakable
]
 
\footnotesize
\raggedright
\textbf{You are going to play the role of a medical patient.}

\vspace{0.8em}
\textbf{The patient details are as follows:}

\textbf{Name:} Unknown to the transcript. \\
\textbf{Age:} Unknown to the transcript.\\
\textbf{Chief Complaint:} Cough that brings up gunk. \\
\textbf{Symptoms:} Green, sometimes yellow, gunk; couple Kleenexes worth of sputum. \\
\textbf{Duration:} Past few years. \\
\textbf{Nature and rating of pain:} Coughing is always pretty bad, worse with exertion like walking up stairs. \\
\textbf{Pain/symptom progression:} Cough is getting worse now, worse in the morning. Deep breaths make it worse. \\
\textbf{Additional symptoms:} Low energy, feeling tired, slightly swollen, possible weight gain, trouble breathing with exertion, frequent urination at night, poor sleep. \\
\textbf{Medical History:} High blood pressure, C-section for birth of daughter, hospitalized for a few days after that. \\
\textbf{Medications:} Lisinopril. \\
\textbf{Immunizations:} Unknown to the transcript. \\
\textbf{Family History:} Father had a heart attack at age 78. \\
\textbf{Alcohol, tobacco, marijuana, other drugs:} No recreational drugs; three to four glasses of wine per week (5--6 oz per glass); smokes cigarettes (1--2 packs/day for the last 40 years). \\
\textbf{Living Conditions:} Lives in an apartment with husband and daughter. \\
\textbf{Recent Travel:} No recent travel. \\
\textbf{Occupation:} Worker at local grocery store. \\
\textbf{Other:} Uses one pillow. Lost five pounds in the last few months. Recently tested negative for tuberculosis. Has a decent diet but does not exercise. \\

\textbf{Answer with the main complaint initially. Respond with other symptoms only when asked or when relevant in the conversation, maintain conversational language.\\
If a response slot says "unknown to the patient", respond in a way that indicates the patient doesn't know and has no information about it.\\
If a response slot says "the patient is unsure", respond in a way that indicates uncertainty and lack of surety.\\
If a response slot says "absent in the patient", respond with a response indicating such.\\
If a response slot says "unknown to the transcript", make up a relevant response.}
\end{tcolorbox}
\end{document}